\documentclass[9pt,conference]{IEEEtran}
\IEEEoverridecommandlockouts
\usepackage{cite}
\usepackage{amsmath,amssymb,amsfonts}
\usepackage{algorithmic}
\usepackage{graphicx}
\usepackage{textcomp}
\usepackage{xcolor}
\def\BibTeX{{\rm B\kern-.05em{\sc i\kern-.025em b}\kern-.08em
    T\kern-.1667em\lower.7ex\hbox{E}\kern-.125emX}}

\usepackage{booktabs}
\usepackage{subcaption}
\usepackage{soul}
\usepackage{url}

\usepackage{tabularx}
\newcolumntype{P}{>{\raggedright\arraybackslash}X}
\newcolumntype{C}{>{\hsize=0.5\hsize}X}
\usepackage{makecell}
\usepackage{multirow}
\usepackage{pifont}

\newcommand{\cmark}{\ding{51}}%
\newcommand{\xmark}{\ding{55}}%

\begin{document}
\bstctlcite{IEEEexample:BSTcontrol}

\title{Gemmini: Enabling Systematic Deep-Learning Architecture Evaluation via Full-Stack Integration}



\author{\IEEEauthorblockN{Hasan Genc*, Seah Kim*, Alon Amid*, Ameer Haj-Ali*, Vighnesh Iyer*, Pranav Prakash*, Jerry Zhao*, Daniel Grubb*,\\ Harrison Liew*, Howard Mao*, Albert Ou*, Colin Schmidt*, Samuel Steffl*, John Wright*, Ion Stoica*,\\ Jonathan Ragan-Kelley†, Krste Asanovic*, Borivoje Nikolic*, Yakun Sophia Shao*}
\IEEEauthorblockA{\textit{*UC Berkeley, †MIT} \\
hngenc@berkeley.edu}
\vspace{-0.25in}
}

\maketitle



\begin{abstract}

DNN accelerators are often developed and evaluated in isolation without considering the cross-stack, system-level effects in real-world environments. This makes it difficult to appreciate the impact of System-on-Chip (SoC) resource contention, OS overheads, and programming-stack inefficiencies on overall performance/energy-efficiency. To address this challenge, we present Gemmini, an open-source\footnote{\url{https://github.com/ucb-bar/gemmini}},
full-stack DNN accelerator generator. Gemmini generates a wide design-space of efficient ASIC accelerators from a flexible architectural template, together with flexible programming stacks and full SoCs with shared resources that capture system-level effects. Gemmini-generated accelerators have also been fabricated, delivering up to three orders-of-magnitude speedups over high-performance CPUs on various DNN benchmarks. 

\end{abstract}

\section{Introduction}

Deep neural networks (DNNs) have gained major interest in recent years in application domains ranging from computer vision, to machine translation, to robotic manipulation.
However, running modern, accurate DNNs with high performance and low energy consumption is often challenging without dedicated accelerators which are difficult and expensive to design.
The demand for cheaper, high-productivity hardware design has motivated a number
of research efforts to develop highly-parameterized and modular hardware
generators for DNN
accelerators and other hardware building blocks~\cite{moreau2018, venkatesan2019magnet,polysa,zhang2018,automated-systolic-cnn-fpgas,deepburning,hybrid-dnn}.
While the hardware generator efforts make it easier to instantiate a DNN accelerator,
they primarily focus on the design of the accelerator component itself, rather than taking into consideration the system-level parameters that determine the overall SoC and the full software stack.
Some industry perspectives have advocated for a more
holistic exploration of DNN accelerator development and deployment~\cite{datacenter-facebook,edge-facebook,ai-tax-hpca}.
However, existing DNN generators have little support for a full-stack programming interface which provides both high and low-level control of the accelerator, and little support for full SoC integration, making it challenging to evaluate system-level implications. 

In this work, we present Gemmini, an open-source, full-stack DNN accelerator
generator for DNN workloads, enabling end-to-end, full-stack implementation and evaluation
of custom hardware accelerator systems for rapidly evolving DNN workloads.
Gemmini's hardware template and parameterization allows users to tune the
hardware design options across a broad spectrum spanning performance, efficiency,
and extensibility.
Unlike existing DNN accelerator generators that focus on standalone accelerators, Gemmini also provides a complete solution spanning both the hardware and software stack, and a complete SoC integration that is compatible with the RISC-V ecosystem.
In addition, Gemmini implements a multi-level software stack with an easy-to-use programming interface to support different programming requirements, as well as tight integration with Linux-capable SoCs which enable the execution of any arbitrary software.


\begin{table*}[t]
\begin{tabularx}{\linewidth}{ c | >{\centering}p{2.5cm} | cccccccc }
\toprule
& Property
& NVDLA
& VTA
& PolySA
& DNNBuilder
& MAGNet
& DNNWeaver
& MAERI
& Gemmini \\
\midrule
\multirow{4}{*}{\makecell{Hardware\\Architecture\\Template}} & Datatypes & Int/Float & Int & Int & Int  & Int & Int & Int & Int/Float \\
& Dataflows & \xmark & \xmark & \cmark & \cmark & \cmark & \cmark & \cmark & \cmark \\ 
& \makecell{Spatial Array} & vector & vector & systolic & systolic & vector & vector & vector & vector/systolic \\
& Direct Convolution & \cmark & \xmark & \xmark & \cmark &  \cmark & \cmark & \cmark & \cmark \\
\hline
\multirow{2}{*}{\makecell{Programming\\Support}} & Software Ecosystem & Compiler & TVM &  SDAccel & Caffe & C & Caffe & Custom & ONNX/C \\ 
& Virtual Memory & \xmark & \xmark & \xmark & \xmark & \xmark & \xmark & \xmark & \cmark \\
\hline
\multirow{2}{*}{\makecell{System\\Support}} & \makecell{Full SoC} & \xmark & \xmark & \xmark & \xmark & \xmark & \xmark & \xmark & \cmark \\
& \makecell{OS Support} & \cmark & \cmark & \xmark & \xmark & \xmark & \xmark & \xmark & \cmark \\ 

\bottomrule
\end{tabularx}
\caption{Comparison of DNN accelerator generators.}
\label{tab:generator-comparison}
\vspace{-0.2in}
\end{table*}


Gemmini-generated accelerators have been successfully fabricated in both TSMC
$16nm$ FinFET and Intel $22nm$ FinFET Low Power (22FFL) process technologies, demonstrating that they can be physically realized.
In addition, our evaluation shows that Gemmini-generated accelerators deliver comparable
performance to a state-of-the-art, commercial DNN
accelerator~\cite{nvdla-hotchips} with a similar set of hardware configurations and achieve up to 2,670x speedup with respect to a baseline CPU.
Gemmini's fully-integrated, full-stack flow enables users to co-design the accelerator, application, and system all at once, opening up new research opportunities for future DL SoC integration. Specifically, in our Gemmini-enabled case studies, we demonstrate how designers can use Gemmini to optimize virtual address translation mechanisms for DNN accelerator workloads, and to partition memory resources in a way that balances the different compute requirements of different layer types within a DNN.


In brief, this work makes the following contributions:

\begin{enumerate}
\item We build Gemmini, an open-source, full-stack DNN accelerator design
infrastructure to enable systematic evaluation of deep-learning architectures.
Specifically, Gemmini provides a flexible hardware template, a multi-layered
software stack, and an integrated SoC environment (Section~\ref{sec:generator}).

\item We perform rigorous evaluation of Gemmini-generated accelerators using FPGA-based performance measurement and commercial ASIC synthesis flows for performance and efficiency analysis. Our evaluation demonstrates that Gemmini-generated accelerators deliver comparable performance compared to state-of-the-art, commercial DNN accelerators (Section~\ref{sec:evaluation}). 

\item We demonstrate that the Gemmini infrastructure enables system-accelerator co-design of SoCs running DNN workloads, including the design of efficient virtual-address translation schemes for DNN accelerators and the provisioning of memory resources in a shared cache hierarchy (Section~\ref{sec:casestudies}).
\end{enumerate}

\begin{figure}[t]
    \centering
    \includegraphics[width=0.9\linewidth]{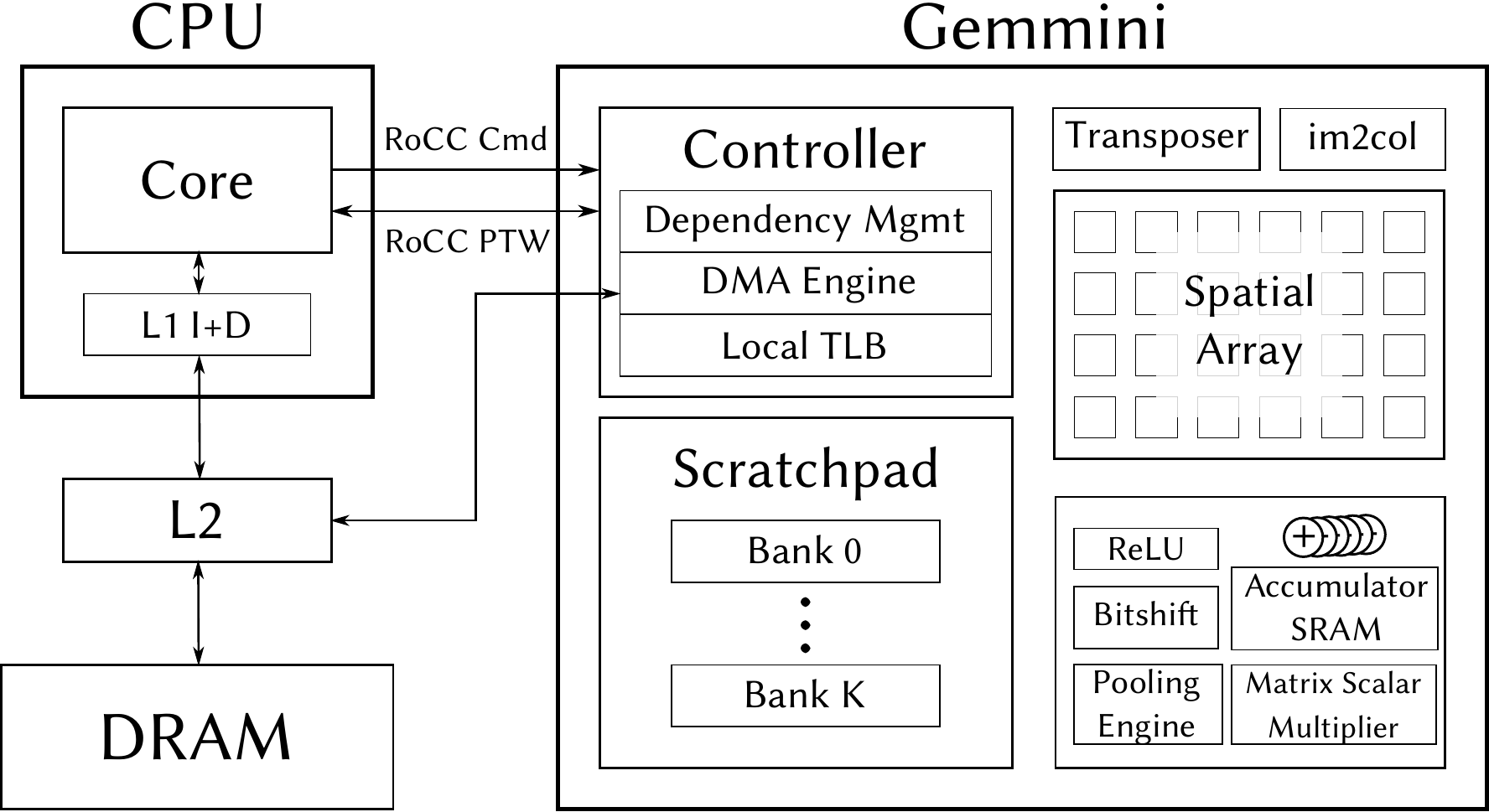}
    \caption{Gemmini hardware architectural template overview.}
    \vspace{-0.3cm}
    \label{fig:arch-template}
    \vspace{-0.1in}
\end{figure}

\section{Background and Motivation}\label{sec:motivation}

The demand for fast and efficient DNN execution from edge to cloud has led to a significant effort in developing novel accelerator instances that are
specialized for different DNN algorithms and/or different deployment scenarios.
This section discusses recent advances in DNN accelerators and
DNN accelerator generators, motivating the need for a full-stack approach to
evaluate deep learning architectures.

\subsection{DNN Accelerators}

Researchers have proposed a large variety of novel DNN accelerators with different performance and energy efficiency targets for different applications across a diverse set of deployment scenarios~\cite{eyeriss2,shidiannao,scaledeep,moreau2018}. At the architecture level, different DNN accelerators exploit different reuse patterns to build specialized memory hierarchies~\cite{interstellar-asplos2020} and interconnect networks~\cite{maeri-asplos2018} to improve performance and energy efficiency. Most existing hardware DNN architectures are largely spatial, where parallel execution units are laid out spatially either in a systolic fashion, as in the case of the TPU, or in parallel vector units like Brainwave~\cite{brainwave-isca-2018} and NVDLA~\cite{nvdla-hotchips}. Based on these architectural templates, recent advances have also started exploring how to leverage applications' sparsity patterns~\cite{sigma-hpca2020,sparse-tpu,sparse-train} and/or emerging in-memory computing technology~\cite{pipelayer,algo-hardware-codesign-for-in-memory}.

\subsection{DNN Accelerator Generators}
Recent research has designed hardware generators 
for DNN accelerators~\cite{nvdla-hotchips,moreau2018,polysa,venkatesan2019magnet,dnnweaver,maeri-asplos2018}.
Table~\ref{tab:generator-comparison} compares the different features supported by existing hardware generators compared to Gemmini.
In contrast to building specific instances of hardware, generator-based
approaches provide parameterizable architectural
templates that can generate a wide variety of hardware and software instances, improving
hardware design productivity.
Here, we discuss the hardware, software, and system level requirements for DNN
accelerator generators to enable full-stack, systematic DNN architecture
evaluation.

DNN accelerator generators must provide flexible architectural templates to cover a wide variety of different DNN accelerator instances, each suited for a different execution environment and a different area/power/performance target. Most DNN accelerator generators today focus only on fixed-point representations and/or only support a single dataflow. In addition, today's generators only target a specific spatial array type, \textit{i.e.}, systolic-based (as in the TPU) or vector-based (as in NVDLA), making it challenging to systematically compare against them. In contrast, Gemmini supports 1) both floating and fixed point data types to handle data representations in training and inference, 2) multiple dataflows that can be configured at design time and run time, 3) both vector and systolic spatial array architectures, enabling quantitative comparison of their efficiency and scalability differences, and 4) direct execution of different DNN operators.
    
Moreover, DNN accelerator generators also need to provide an easy-to-use
programming interface so that end users can
quickly program their applications for the generated accelerators.
Different developers would prefer different software design environments based upon their targets or research interests.
For example, DNN application practitioners would prefer that the hardware
programming environment be hidden by DNN development frameworks like
PyTorch or TVM so that they don't need to worry about low-level development
details, as in the case of VTA~\cite{moreau2018} and DNNWeaver~\cite{dnnweaver}.
At the same time, framework developers and system programmers may want to
interact with the hardware at a low level, in either C/C++ or assembly, to
accurately control hardware states and squeeze every bit of efficiency out, as in the case of MAGNet~\cite{venkatesan2019magnet} and Maeri~\cite{maeri-asplos2018}.
Unlike other DNN generators that tend to focus on one of these
requirements, Gemmini provides a multi-level programming interface
to satisfy users with different requirements.
In addition, Gemmini is the first infrastructure that provides hardware support for virtual memory without the need for any special driver software, making it significantly easier for end-users to program accelerators.

Third, system-level integration, including both the SoC and the system software, is also critical in DNN accelerator generators.
Today's DNN accelerators are typically designed and evaluated in isolation. However, when they are eventually deployed, they need to be integrated as part of a larger system.
In fact, recent industry evaluations have demonstrated that modern ML
workloads could spend as much as 77\% of their time running on CPUs, even in the
presence of a hardware accelerator, to execute either new operators or to move
data between the CPU and accelerators~\cite{centaur-isca2020, edge-facebook,datacenter-facebook,ai-tax-hpca}.
However, unfortunately, none of the existing DNN accelerator generators support full SoC integration with host CPUs and shared resources like caches and system buses. Motivated by this observation, Gemmini has built-in system integration support where users can directly instantiate a complete SoC environment that can boot Linux, directly enabling architects to evaluate subtle trade-offs at the
system level.

\vspace{-0.01in}
\section{Gemmini Generator}\label{sec:generator}


Gemmini is an open-source, full-stack generator of DNN accelerators, spanning across different hardware architectures, programming interfaces, and system integration options.
With Gemmini, users can generate everything from low-power edge accelerators to high-performance cloud accelerators equipped with out-of-order CPUs. Users can then investigate how the hardware, SoC, OS, and software overhead interact to affect overall performance and efficiency.

\subsection{Architectural Template}
\label{ssec:arch}

\begin{figure}[t]
    \centering
    \includegraphics[width=0.85\linewidth]{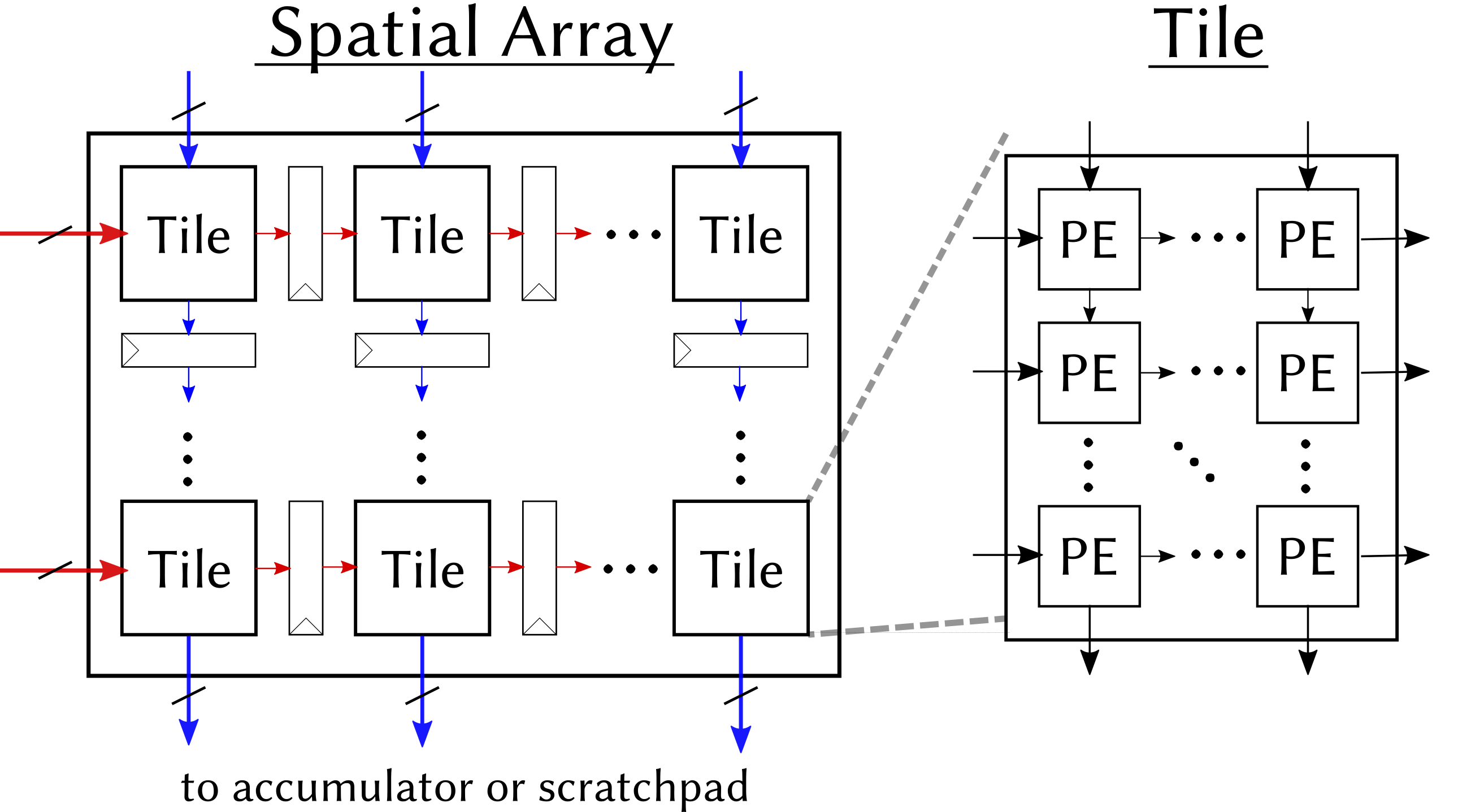}
    \caption{Microarchitecture of Gemmini's two-level spatial array.} 
    \label{fig:systolic_arch}
    \vspace{-0.2in}
\end{figure}

Figure~\ref{fig:arch-template} illustrates Gemmini's architectural template.
The central unit in Gemmini's architectural template is a spatial architecture with spatially distributed processing elements (PEs), each of which performs dot products and accumulations.
The spatial array reads data from a local, explicitly managed scratchpad of banked SRAMs, while it writes results to a local accumulator storage with a higher bitwidth than the inputs.
Gemmini also supports other commonly-used DNN kernels, \textit{e.g.}, pooling, non-linear activations (ReLU or ReLU6), and matrix-scalar multiplications, through a set of configurable, peripheral circuitry.
Gemmini-generated accelerators can also be integrated with a RISC-V host CPU
to program and configure accelerators.


We design Gemmini's spatial array with a two-level hierarchy to provide a flexible template for different microarchitecture structures, as demonstrated in Figure~\ref{fig:systolic_arch}.
The spatial array is first composed of \textit{tiles}, where tiles are connected via explicit pipeline registers.
Each of the individual tiles can be further broken down into an array of PEs, where PEs in the same tile are connected combinationally without pipeline registers.
Each PE performs a single multiply-accumulate (MAC) operation every cycle, using either the weight- or the output-stationary dataflow.
The tiles are composed of rectangular arrays of PEs, where PEs in the same tile are connected combinationally with no pipeline registers in between them. The spatial array, likewise, is composed of a rectangular array of tiles, but each tile \textit{does} have pipeline registers between it and its neighbors.
Every PE and every tile shares inputs and outputs only with its adjacent neighbors.

\begin{figure}[t]
\centering
\includegraphics[width=\linewidth]{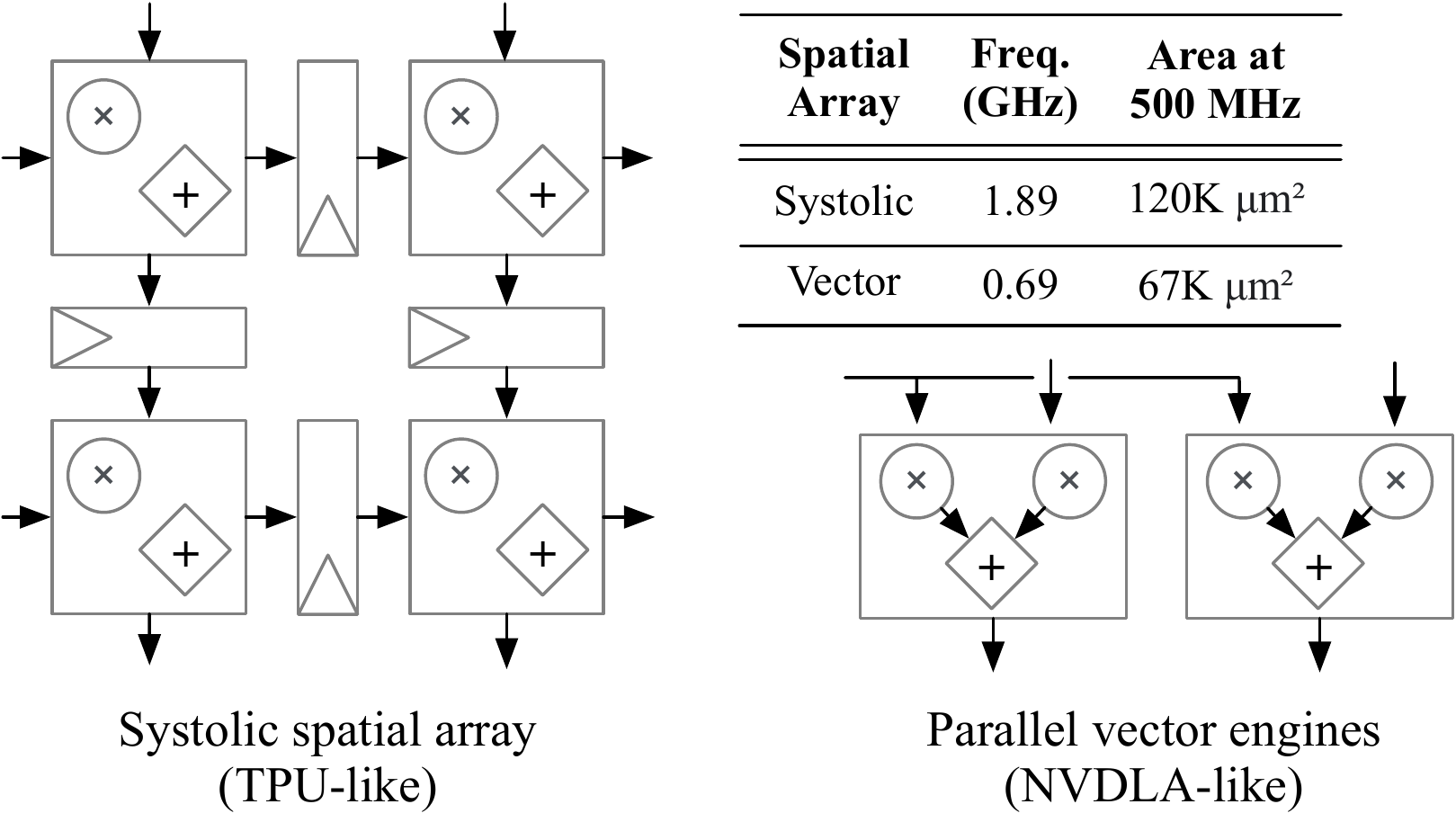}
\vspace{-0.2in}
\caption{Examples of two different spatial architectures generated by Gemmini. Both perform four multiply-accumulates per cycle though with different connectivities between multiply-and-accumulate units.}
\label{fig:nvdla-vs-tpu-comparison}
\vspace{-0.2in}
\end{figure}

Figure~\ref{fig:nvdla-vs-tpu-comparison} illustrates how Gemmini's two-level hierarchy provides the flexibility to support anything from fully-pipelined TPU-like architectures to NVDLA-like parallel vector engines where PEs are combinationally joined together to form MAC reduction trees, or any other design points in between these two extremes. We synthesized both designs with 256 PEs. We found that the TPU-like design achieves a 2.7x higher maximum frequency, due to its shorter MAC chains, but consumes 1.8x as much area as the NVDLA-like design, and 3.0x as much power, due to its pipeline registers. With Gemmini, designers can explore such footprint vs. scalability trade-offs across different accelerator designs.

\subsection{Programming Support}
\label{software}

The Gemmini generator produces not just a hardware stack, but also a tuned software stack, boosting developers' productivity as they explore different hardware instantiations.
Specifically, Gemmini provides a multi-level software flow to support different programming scenarios.
At the \textit{high level}, Gemmini contains a push-button software flow which reads DNN descriptions in the ONNX file format
and generates software binaries that will run them, mapping as many kernels as possible onto the Gemmini-generated accelerator. Alternatively, at the \textit{low level}, the generated accelerator can also be programmed through C/C++ APIs, with tuned functions for common DNN kernels. These functions must be tuned differently for different hardware instantiations in order to achieve high performance, based on scratchpad sizes and other parameters. Therefore, every time a new accelerator is produced, Gemmini also generates an accompanying header file containing various parameters, \textit{e.g.} the dimensions of the spatial array, the dataflows supported, and the compute blocks that are included (such as pooling, im2col, or transposition blocks).

\begin{figure}[t]
    \centering
    \includegraphics[width=\linewidth]{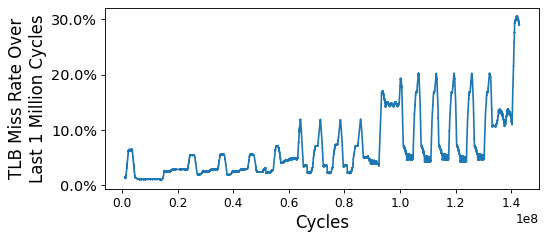}
    \caption{TLB miss rate over a full ResNet50 inference, profiled on a Gemmini-generated accelerator.}
    \label{fig:tlb_hit_rate}
    \vspace{-0.2in}
\end{figure}

\textbf{Data Staging and Mapping:}
At runtime, based on the dimensions of a layer's inputs, and the hardware parameters of the accelerator instantiation, Gemmini uses heuristics to maximize the amount of data moved into the scratchpad per iteration.
Gemmini calculates loop tile sizes at runtime, and these tile sizes determine when and how much data is moved between DRAM, L2, and scratchpad during the execution of our tiled matrix multiplication, convolution, residual-addition, etc. kernels.
If the programmer wishes, the low-level API also allows them to manually set tile-sizes for each kernel.

\textbf{Virtual Memory Support:}
In addition to the programming interface, Gemmini also makes it easier to program accelerators by providing virtual memory support. This is useful for programmers who wish to avoid manual address translations as well as for researchers who wish to investigate virtual memory support in modern accelerators.
Gemmini also enables users to co-design and profile their own virtual address translation system. For example, Figure~\ref{fig:tlb_hit_rate} shows the miss rate of an example accelerator's local TLB profiled on Gemmini. As we can see, the miss rate occasionally climbs to 20-30\% of recent requests, due to the tiled nature of DNN workloads, which is orders-of-magnitude greater than the TLB miss rates recorded in prior CPU non-DNN benchmarks~\cite{lustig2013tlb}.
Later, in Section~\ref{virtual-memory-case-study}, we use Gemmini to co-design a virtual address translation system which achieves near-maximum end-to-end performance on accelerated DNN workloads, with only a few TLB entries in total.

\subsection{System Support}
\label{system}



\begin{figure}[t]
     \centering
     \includegraphics[width=0.75\linewidth]{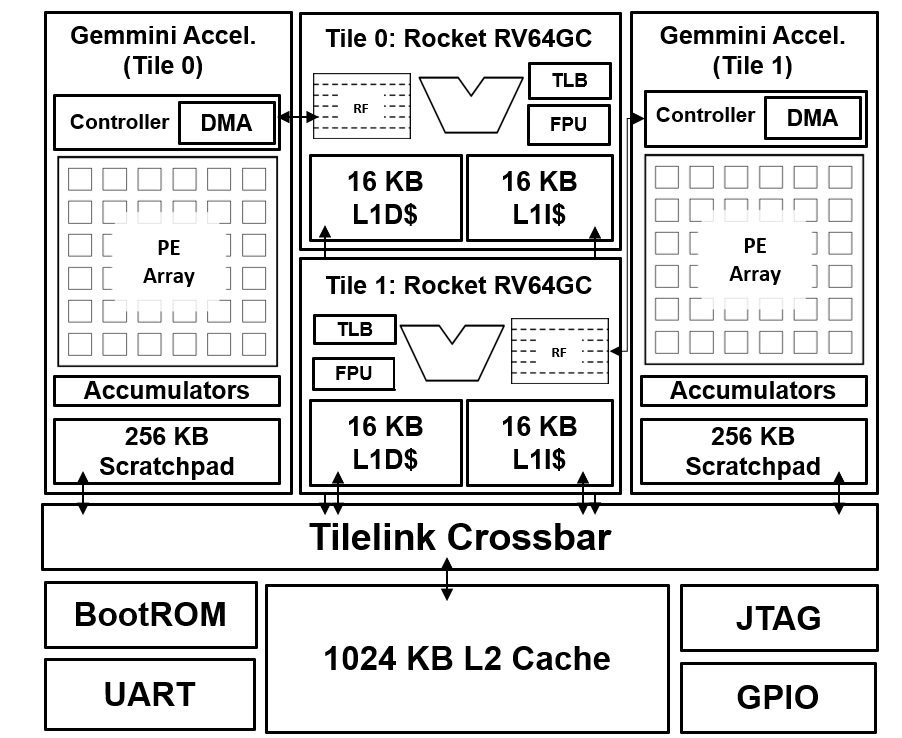}
     \caption{Example dual-core SoC with a Gemmini accelerator attached to each CPU, as well as a shared L2 cache and standard peripherals.}
     \label{fig:contention-soc}
     \vspace{-0.2in}
 \end{figure}

Gemmini allows architects to integrate RISC-V CPUs 
with Gemmini-generated accelerators in the Chipyard~\cite{chipyard} framework.
These can range from simple, in-order microcontrollers which are not expected to do much more than IO management, all the way up to out-of-order, high-performance, server-class CPUs that may be running multiple compute-intensive applications even as they are sending commands to the Gemmini-generated accelerator.
SoCs can also be configured to host \textit{multiple} host CPUs and Gemmini-generated accelerators, which can each operate on different tasks in parallel with each other. Figure~\ref{fig:contention-soc} is one example of a dual-core system, where each CPU has its own Gemmini-generated accelerator.
Additional SoC-level parameters include bus widths between accelerators and host CPUs, as well as the size, associativity and hierarchy of the caches in the multicore, multicache memory system. Later, in Section~\ref{cache-contention}, we show how these parameters can be tuned, based on the computational characteristics of DNNs, to improve performance by over~8\%.

RISC-V-based full SoC integration also enables deep software-stack support, such that Gemmini-generated accelerators can easily be evaluated running the full software stack up to and including the operating system itself. This enables early exploration of accelerated workloads in a realistic environment where context switches, page table evictions, and other unexpected events can happen at any time. These unexpected events can uncover bugs and inefficiencies that a ``baremetal'' environment would not bring to the surface. For example, our experience of running Linux while offloading DNN kernels to a Gemmini-generated accelerator uncovered a non-deterministic deadlock that would only occur if context switches happened at very particular, inopportune times.
Running on a full software stack with an OS also uncovered certain bugs where Gemmini read from certain regions of physical memory without the proper permissions. On a ``baremetal'' environment, these violations were silently ignored.

\section{Gemmini Evaluation}\label{sec:evaluation}
This section discusses our evaluation methodology and evaluation results of Gemmini-generated accelerators compared to both CPUs and state-of-the-art, commercial accelerators.

\subsection{Evaluation Methodology}

We evaluate the end-to-end performance of Gemmini-generated accelerators using the FireSim FPGA-accelerated simulation platform~\cite{karandikar2018firesim}. We evaluate five popular DNNs: ResNet50, AlexNet, SqueezeNet v1.1, MobileNetV2, and BERT.
All DNNs are evaluated with a full Linux environment on a complete cycle-exact simulated SoC.
We synthesize designs using Cadence Genus with the Intel 22nm FFL process technology and place-and-route them using Cadence Innovus.
Our layout and area breakdown, described in Figure~\ref{fig:evaluation}, show that the SRAMs alone consume 67.1\% of the accelerator's total area. The spatial array itself only consumes 11.3\%, while the host CPU consumed a higher 16.6\% of area.


\subsection{Performance Results}

We evaluated the performance of several Gemmini configurations, with different host CPUs and different ``optional'' compute blocks, to determine how the accelerator and host CPU configuration may interact to impact end-to-end performance. In particular, we evaluated two host CPUs: a low-power in-order Rocket core, and a high-performance out-of-order BOOM core. We used two different Gemmini configurations:
one \textit{without} an optional im2col block, and the other \textit{with} an im2col block which allowed the accelerator to perform im2col on-the-fly, relieving the host CPU of that burden.

\begin{figure}[b]
\centering
\includegraphics[width=\linewidth]{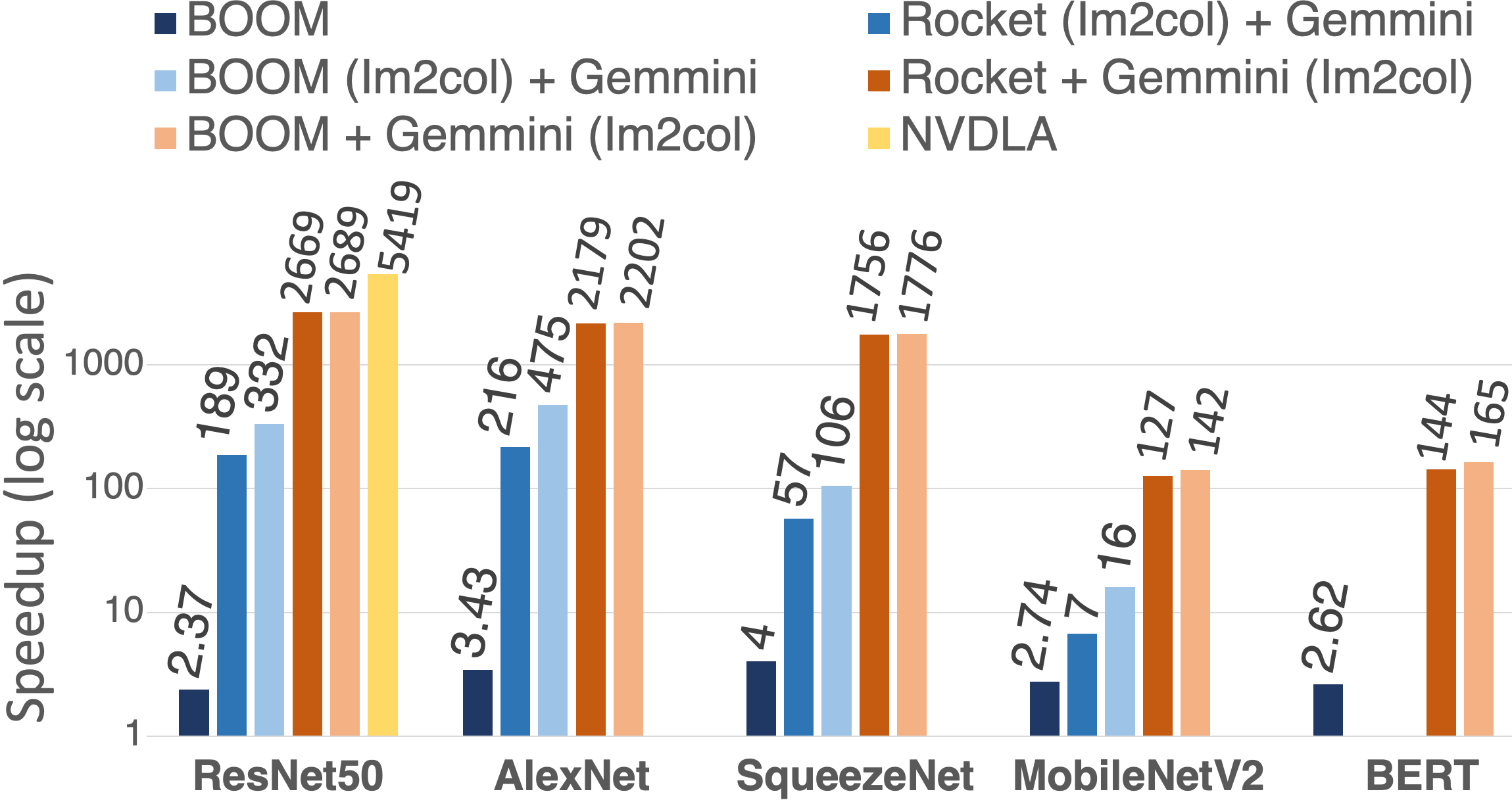}
\caption{Speedup compared to an in-order CPU baseline. For CNNs, im2col was performed on either the CPU, or on the accelerator.}
\label{fig:perf-dnn}
\end{figure}

As illustrated in Figure~\ref{fig:perf-dnn}, when the accelerator is built without an on-the-fly im2col unit, its performance depends heavily on the host-CPU which becomes responsible for performing im2col during CNN inference.
A larger out-of-order BOOM host CPU increases performance by 2.0x across all CNNs.
The less complex the DNN accelerator is, the more the computational burden is shifted onto the CPU, giving the host CPU a larger impact on end-to-end performance.

\begin{figure}[t]
\begin{subfigure}[b]{0.5\linewidth}
\centering
\scalebox{0.8} {
\begin{tabular}{ l | r | r }
\hline
\textbf{Component size} &  \textbf{\makecell{Area\\($\mu m ^2$)}} & \textbf{\makecell{\% of \\System\\Area}} \\
\hline
\hline
Spatial Array (16x16) & 116K & 11.3\% \\
Scratchpad (256 KB) & 544K & 52.9\%  \\
Accumulator (64 KB) & 146K & 14.2\%  \\
CPU (Rocket, 1 core) & 171K & 16.6\% \\
\hline
Total & 1,029K & 100.0\% \\
\hline
\end{tabular}
}
\caption{Area breakdown.}
\label{tab:area-table}
\end{subfigure}
\hfill
\begin{subfigure}[b]{0.4\linewidth}
\centering
\includegraphics[width=0.8\linewidth]{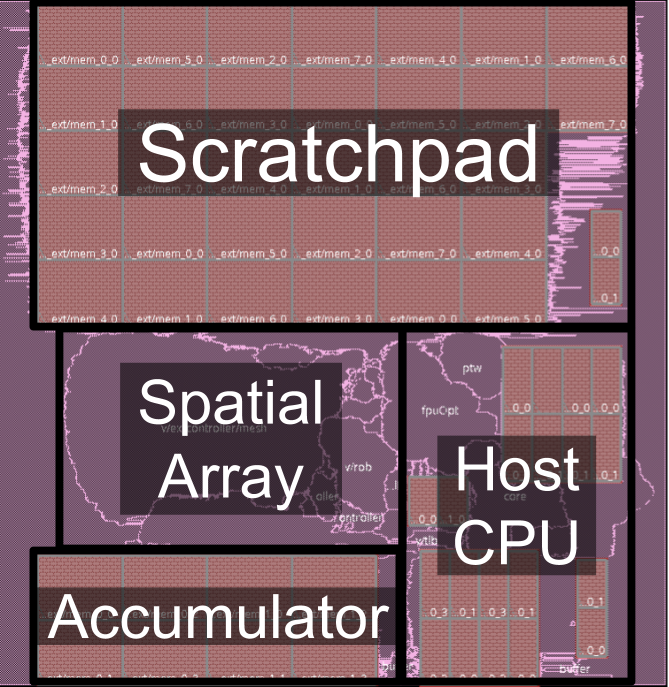}
\caption{Layout.}
\label{fig:area-layout}
\end{subfigure}
\caption{Area breakdown and layout of accelerator with host CPU.}
\label{fig:evaluation}
\vspace{-0.2in}
\end{figure}

However, when the accelerator is equipped with an on-the-fly im2col unit, the choice of host CPU is far less important, because the CPU's computational burden is shifted further onto the accelerator. Adding a small amount of complexity to the accelerator allows us to reduce the area and complexity of the host CPU to a simple in-order core while preserving performance. Gemmini enables hardware designers to easily make these performance-efficiency tradeoffs.

With the on-the-fly im2col unit and a simple in-order Rocket CPU, Gemmini achieves 22.8 frames per second (FPS) for ResNet50 inference when running at 1 GHz, which is a 2,670x speedup over the in-order Rocket CPU and an 1,130x speedup over the out-of-order BOOM CPU. The accelerator also achieves 79.3 FPS on AlexNet. Some DNN models such as MobileNet are not efficiently mapped to spatial accelerators due to the low data reuse within the depthwise convolution layers. Therefore, Gemmini demonstrates only a 127x speedup compared to the Rocket host CPU on MobileNetV2, reaching 18.7 FPS at 1GHz.  On SqueezeNet, which was designed to be run efficiently on modern CPUs while conserving memory bandwidth, Gemmini still demonstrates a 1,760x speedup over the Rocket host CPU. Our results are comparable to other accelerators, such as NVDLA, when running with the same number of PEs as the configuration in Figure~\ref{tab:area-table}. When running language models such as BERT, Gemmini achieves a 144x improvement over the Rocket CPU.

\section{Gemmini Case Studies}\label{sec:casestudies}
This section demonstrates how Gemmini enables full system co-design with two case studies. We use Gemmini to design a novel virtual address translation scheme, and to find the optimal SoC-level resource partition scheme of a multi-core, multi-accelerator system.

\subsection{Virtual Address Translation}
\label{virtual-memory-case-study}

With an RTL-level implementation that supports virtual memory, users can co-design their own virtual address translation schemes based on their accelerator and SoC configuration. Prior works in virtual address translation for DNN accelerators have proposed very different translation schemes, from NeuMMU~\cite{neummu-asplos2020}, which calls for a highly parallel address-translation system with 128 page-table walkers (PTWs), to Cong et al.~\cite{address-trans-cong-hpca2017}, who recommend a more modest two-level TLB hierarchy, with the host CPU's default PTW co-opted to serve requests by the accelerator. This lack of convergence in the prior literature motivates a platform that allows co-design and design-space exploration of the accelerator SoC together with its virtual address translation system, for both hardware designers and researchers. Fortunately, with Gemmini, we can iterate over a variety of address translation schemes as we tune the accelerator and SoC.

\begin{figure}
    \centering
    \begin{subfigure}[t]{0.439\linewidth}
        \centering
        \includegraphics[width=\linewidth]{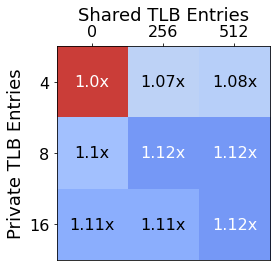}
        \caption{Without filter registers.}
        \label{fig:tlb_wait_cycles}
    \end{subfigure}
    \hfill
    \begin{subfigure}[t]{0.439\linewidth}
        \centering
        \includegraphics[width=\linewidth]{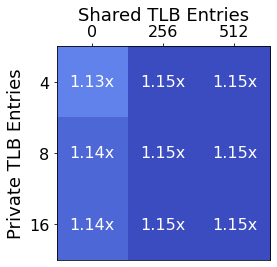}
        \caption{With filter registers.}
        \label{fig:tlb_wait_cycles_with_filter}
    \end{subfigure}
    \caption{Normalized performance of ResNet50 inference on Gemmini-generated accelerator with different private and shared TLB sizes.}
    \vspace{-0.2in}
\end{figure}

To demonstrate, we configure Gemmini to produce a two-level TLB cache, with one private TLB for the accelerator, and one larger shared TLB at the L2 cache that the private TLB falls back on when it misses. Our design includes only one PTW, shared by both the CPU and the accelerator, which is suitable for low-power devices. We configure the accelerator for low-power edge devices, with a 16-by-16 systolic mesh and a 256 KB scratchpad. As shown in Figure~\ref{fig:tlb_wait_cycles}, we iterate over a variety of TLB sizes to find the design that best balances TLB overhead and overall performance, including over a design point where the shared L2 TLB has zero entries.

Figure~\ref{fig:tlb_wait_cycles} demonstrates that the private accelerator TLB has a far greater impact on end-to-end performance than the much larger shared L2 TLB. Increasing the private TLB size from just four to 16 improves performance by up to 11\%. However, adding even 512 entries to the L2 TLB never improves performance by more than 8\%. This is because our workloads exhibit high page locality; even with tiled workloads, our private TLB's hit rate remained above 84\%, even with the smallest TLB sizes we evaluated. In fact, we found that 87\% of consecutive \textit{read} TLB requests, and 83\% of consecutive \textit{write} TLB requests, were made to the same page number, demonstrating high page locality. However, because reads and writes were overlapped, read and write operations could evict each other's recent TLB entries.

Although tuning TLB sizes improves hit rates, our private TLB hit latency in the tests shown in Figure~\ref{fig:tlb_wait_cycles} was still several cycles long. Fortunately, using the Gemmini platform, we were able to implement a simple optimization: a single register that caches the last TLB hit for read operations, and another register that caches TLB hits for write operations. These two registers allow the DMA to ``skip'' the TLB request if two consecutive requests are made to the same virtual page number, and help reduce the possibility of read-write contention over the TLB. These ``filter registers'' reduce the TLB hit latency to 0 cycles for consecutive accesses to the same page. As Figure~\ref{fig:tlb_wait_cycles_with_filter} shows, this low-cost optimization significantly improves our end-to-end performance, especially for small private TLB sizes. Due to our high TLB hit rate and low TLB hit penalty, we found that a very small 4-entry private TLB equipped with filter registers, but without an expensive shared L2 TLB, achieved only 2\% less than the maximum performance recorded. With such a configuration, the private TLB hit rate (including hits on the filter registers) reached 90\% and further increases to either TLB's size improved performance by less than 2\%, even if hundreds of new TLB entries were added.

Using Gemmini, we have demonstrated that a modest virtual address translation system, with very small private TLBs, a single page-table-walker, and two low-cost filter registers for the TLB, can achieve near maximum performance for low-power edge devices. Gemmini is designed to enable such co-design of the SoC and its various components, such as its virtual address translation system.

\subsection{System-Level Resource Partition}
\label{cache-contention}

Gemmini also enables application-system co-design for real-world DNN workloads. To demonstrate, we present a case study describing a system-level design decision: memory partitioning based on application characteristics. We investigate memory partitioning strategies in both single-core and multi-core SoCs.


\begin{figure*}[t]
\centering
\begin{subfigure}[b]{0.325\textwidth}
    \centering
    \scalebox{0.875} {
    \begin{tabular}{ l | c | c | c }
    \hline
    \makecell{Config\\Name} & \makecell{Scratchpad\\(per core)} & \makecell{Accumulator\\(per core)} & \makecell{L2\\Cache} \\  
    \hline
    \hline
    Base & 256 KB & 256 KB & 1 MB   \\
    BigSP & 512 KB & 512 KB & 1 MB  \\
    BigL2 & 256 KB & 256 KB & 2 MB  \\ \hline
    \multicolumn{1}{c}{} & \multicolumn{1}{c}{} & \multicolumn{1}{c}{} & \multicolumn{1}{c}{} \\
    \end{tabular}
    }
    \caption{Resource contention SoC configurations}
    \label{tab:contention-soc-configs}
\end{subfigure}
\hfill
\begin{subfigure}[b]{.325\textwidth}
    \vspace{0pt}
      \centering
      \includegraphics[width=\linewidth]{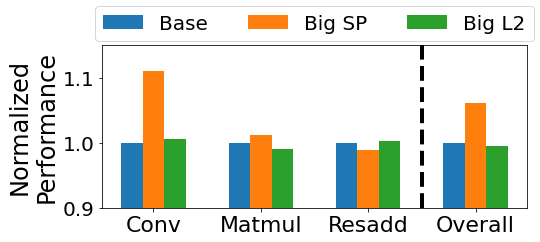}
      \caption{Performance of single-core SoCs.}
      \label{fig:single-contention}
\end{subfigure}
\hfill
\begin{subfigure}[b]{.325\textwidth}
    \vspace{0pt}
      \centering
      \includegraphics[width=\linewidth]{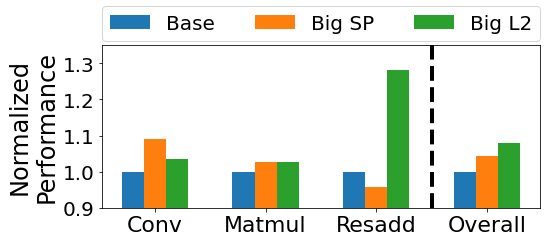}
      \caption{Performance of dual-core SoCs.}
      \label{fig:dual-contention}
\end{subfigure}
    \caption{Performance of the various SoC configurations in the case study, normalized to the performance of the Base configuration.}
    \label{fig:contention}
    \vspace{-0.2in}
\end{figure*}

Real-world DNN applications, such as CNN inference, have diverse layer types which have different computational requirements and which contend for resources on an SoC in different ways. For example, ResNet50 includes convolutions, matrix multiplications, and residual additions, which all exhibit quite different computational patterns. Convolutions have high arithmetic intensity; matrix multiplications have less; and residual additions have almost no data re-use at all. Additionally, unlike the other two types of layers, residual additions benefit most if layer outputs can be stored inside the cache hierarchy for a long time, rather than being evicted by intermediate layers, before finally being consumed several layers later. These different layer characteristics suggest different ideal SoC configurations. To run with optimal performance over an entire DNN, a hardware designer must balance all these constraints.



To demonstrate, we run ResNet50 inference on six different SoC configurations. These are the three different configurations described in Figure~\ref{tab:contention-soc-configs}, repeated for both single- and dual-core SoCs (as in Figure~\ref{fig:contention-soc}), where each CPU core has its own Gemmini-generated accelerator. The dual-core SoCs run two ResNet50 workloads in parallel, while the single-core SoCs run just one. The base design point has a 256 KB scratchpad, and a 256 KB accumulator per core, as well as a 1 MB shared L2 cache. The scratchpad and accumulator memories are private to the accelerators, but the L2 cache is shared by all CPUs and accelerators on the SoC. We presume that we have 1 MB of extra SRAM that we can allocate to our memory system, but we need to decide whether to allocate these SRAMs to the accelerators' private memory, or to the L2 caches.


As shown in Figures~\ref{fig:single-contention} and~\ref{fig:dual-contention}, convolutional layers benefit from a larger, explicitly managed scratchpad, due to their very high arithmetic intensity. Convolutional kernels exhibit a 10\% speedup with one core, and an 8\% speedup in the dual-core case, when the scratchpad and accumulator memory is doubled by the addition of our 1 MB worth of SRAMs. The matmul layers, on the other hand, achieve only a 1\% and 3\% speedup when the scratchpad is enlarged in the single-core and dual-core cases respectively, due to their lower arithmetic intensity. Residual additions, which have virtually no data re-use and are memory-bound operations, exhibit no speedup when increasing the scratchpad memory size. Instead, they exhibit a minor 1\%-4\% slowdown, due to increased cache thrashing. In the single-core case, the increased convolutional and matrix multiplication performance is enough to make the design point with increased scratchpad memory, rather than increased L2 memory, the most performant design point.


However, Figure~\ref{fig:dual-contention} shows that when we run dual-process applications that compete for the same shared L2 cache, allocating the extra 1~MB of memory to the shared L2 cache improves overall performance more than adding that memory to the accelerators' scratchpad and accumulator memories. Increasing the scratchpad size still improves convolutional performance more than increasing the L2 size, but this improvement in performance is more than negated by the 22\% speedup of residual additions that the dual-core BigL2 design point enjoys. This is because each core's residual addition evicts the input layer that the other one is expecting from the shared L2 cache, increasing the latency of memory-bound residual addition layers. The dual-core BigL2 configuration, which increases the shared cache sizes, alleviates this contention, reducing the L2 miss rate by 7.1\% over the full ResNet50 run, and increasing overall performance by 8.0\%. The BigSP configuration, on the other hand, improves overall performance by only 4.2\% in the dual-core case.


With Gemmini, we have demonstrated how the memory partitioning strategy, a key component of system-level design, can be decided based upon application characteristics, such as the composition of layer types and the number of simultaneous running processes.

\section{Conclusion}
We present Gemmini, a full-stack, open-source generator of DNN accelerators that
enables systematic evaluations of DNN accelerator architectures.
Gemmini leverages a flexible architectural template to capture
different flavors of DNN accelerator architectures.
In addition, Gemmini provides a push-button, high-level software flow to
boost programmers' productivity.
Finally, Gemmini generates a full SoC that runs real-world software stacks
including operating systems, to enable system architects to evaluate
system-level impacts. 
Our evaluation shows that Gemmini-generated accelerators demonstrate high performance efficiency, and  
our case studies show how accelerator designers and system architects can
use Gemmini to co-design and evaluate system-level behavior in emerging applications.

\section{Acknowledgements}

This research was, in part, funded by the U.S. Government under the DARPA RTML program (contract FA8650-20-2-7006). The views and conclusions contained in this document are those of the authors and should not be interpreted as representing the official policies, either expressed or implied, of the U.S. Government.



\bibliographystyle{IEEEtran}
\bibliography{ref,sophia}

\begin{thebibliography}{10}
\providecommand{\url}[1]{#1}
\csname url@samestyle\endcsname
\providecommand{\newblock}{\relax}
\providecommand{\bibinfo}[2]{#2}
\providecommand{\BIBentrySTDinterwordspacing}{\spaceskip=0pt\relax}
\providecommand{\BIBentryALTinterwordstretchfactor}{4}
\providecommand{\BIBentryALTinterwordspacing}{\spaceskip=\fontdimen2\font plus
\BIBentryALTinterwordstretchfactor\fontdimen3\font minus
  \fontdimen4\font\relax}
\providecommand{\BIBforeignlanguage}[2]{{%
\expandafter\ifx\csname l@#1\endcsname\relax
\typeout{** WARNING: IEEEtran.bst: No hyphenation pattern has been}%
\typeout{** loaded for the language `#1'. Using the pattern for}%
\typeout{** the default language instead.}%
\else
\language=\csname l@#1\endcsname
\fi
#2}}
\providecommand{\BIBdecl}{\relax}
\BIBdecl

\bibitem{moreau2018}
T.~Moreau \emph{et~al.}, ``{VTA: An Open Hardware-Software Stack for Deep
  Learning},'' \emph{CoRR}, 2018.

\bibitem{venkatesan2019magnet}
R.~Venkatesan \emph{et~al.}, ``{MAGNet: A Modular Accelerator Generator for
  Neural Networks},'' in \emph{ICCAD}, 2019.

\bibitem{polysa}
J.~Cong \emph{et~al.}, ``{PolySA: polyhedral-based systolic array
  auto-compilation},'' in \emph{ICCAD}, 2018.

\bibitem{zhang2018}
X.~Zhang \emph{et~al.}, ``{DNNBuilder: An Automated Tool for Building
  High-performance DNN Hardware Accelerators for FPGAs},'' in \emph{ICCAD},
  2018.

\bibitem{automated-systolic-cnn-fpgas}
{Xuechao Wei} \emph{et~al.}, ``Automated systolic array architecture synthesis
  for high throughput cnn inference on fpgas,'' in \emph{DAC}, 2017.

\bibitem{deepburning}
Y.~{Wang} \emph{et~al.}, ``Deepburning: Automatic generation of fpga-based
  learning accelerators for the neural network family,'' in \emph{DAC}, 2016.

\bibitem{hybrid-dnn}
H.~{Ye} \emph{et~al.}, ``{HybridDNN}: A framework for high-performance hybrid
  dnn accelerator design and implementation,'' in \emph{DAC}, 2020.

\bibitem{datacenter-facebook}
K.~Hazelwood \emph{et~al.}, ``{Applied Machine Learning at Facebook: A
  Datacenter Infrastructure Perspective},'' in \emph{HPCA}, 2018.

\bibitem{edge-facebook}
C.-J. Wu \emph{et~al.}, ``{Machine Learning at Facebook: Understanding
  Inference at the Edge},'' in \emph{HPCA}, 2019.

\bibitem{ai-tax-hpca}
D.~{Richins} \emph{et~al.}, ``{Missing the Forest for the Trees: End-to-End AI
  Application Performance in Edge Data Centers},'' in \emph{HPCA}, 2020.

\bibitem{nvdla-hotchips}
F.~Sijstermans, ``{The NVIDIA Deep Learning Accelerator},'' in \emph{Hot
  Chips}, 2018.

\bibitem{eyeriss2}
Y.~{Chen} \emph{et~al.}, ``{Eyeriss v2: A Flexible Accelerator for Emerging
  Deep Neural Networks on Mobile Devices},'' \emph{JETCAS}, 2019.

\bibitem{shidiannao}
Z.~{Du} \emph{et~al.}, ``{ShiDianNao: Shifting vision processing closer to the
  sensor},'' in \emph{ISCA}, 2015.

\bibitem{scaledeep}
S.~Venkataramani \emph{et~al.}, ``{ScaleDeep: A Scalable Compute Architecture
  for Learning and Evaluating Deep Networks},'' in \emph{ISCA}, 2017.

\bibitem{interstellar-asplos2020}
X.~Yang \emph{et~al.}, ``Interstellar: Using {Halide}’s scheduling language
  to analyze {DNN} accelerators,'' in \emph{ASPLOS}, 2020.

\bibitem{maeri-asplos2018}
H.~Kwon \emph{et~al.}, ``{MAERI: Enabling Flexible Dataflow Mapping over {DNN}
  Accelerators via Programmable Interconnects},'' in \emph{ASPLOS}, 2018.

\bibitem{brainwave-isca-2018}
J.~Fowers \emph{et~al.}, ``{A Configurable Cloud-Scale DNN Processor for
  Real-Time AI},'' in \emph{ISCA}, 2018.

\bibitem{sigma-hpca2020}
E.~{Qin} \emph{et~al.}, ``Sigma: A sparse and irregular gemm accelerator with
  flexible interconnects for dnn training,'' in \emph{HPCA}, 2020.

\bibitem{sparse-tpu}
X.~He \emph{et~al.}, ``Sparse-{TPU}: Adapting systolic arrays for sparse
  matrices,'' in \emph{ICS}, 2020.

\bibitem{sparse-train}
P.~{Dai} \emph{et~al.}, ``{SparseTrain}: Exploiting dataflow sparsity for
  efficient convolutional neural networks training,'' in \emph{DAC}, 2020.

\bibitem{pipelayer}
L.~Song \emph{et~al.}, ``{PipeLayer: A Pipelined ReRAM-Based Accelerator for
  Deep Learning},'' in \emph{HPCA}, 2017.

\bibitem{algo-hardware-codesign-for-in-memory}
H.~{Kim} \emph{et~al.}, ``Algorithm/hardware co-design for in-memory neural
  network computing with minimal peripheral circuit overhead,'' in \emph{DAC},
  2020.

\bibitem{dnnweaver}
H.~Sharma \emph{et~al.}, ``{From High-level Deep Neural Models to FPGAs},'' in
  \emph{MICRO}, 2016.

\bibitem{centaur-isca2020}
G.~{Henry} \emph{et~al.}, ``{High-Performance Deep-Learning Coprocessor
  Integrated into x86 SoC with Server-Class CPUs Industrial Product},'' in
  \emph{ISCA}, 2020.

\bibitem{lustig2013tlb}
D.~Lustig \emph{et~al.}, ``Tlb improvements for chip multiprocessors:
  Inter-core cooperative prefetchers and shared last-level tlbs,'' \emph{TACO},
  2013.

\bibitem{chipyard}
A.~Amid \emph{et~al.}, ``{Chipyard: Integrated Design, Simulation, and
  Implementation Framework for Custom SoCs},'' \emph{IEEE Micro}, 2020.

\bibitem{karandikar2018firesim}
S.~{Karandikar} \emph{et~al.}, ``{FireSim: FPGA-Accelerated Cycle-Exact
  Scale-Out System Simulation in the Public Cloud},'' in \emph{ISCA}, 2018.

\bibitem{neummu-asplos2020}
B.~Hyun \emph{et~al.}, ``{NeuMMU: Architectural Support for Efficient Address
  Translations in Neural Processing Units},'' in \emph{ASPLOS}, 2020.

\bibitem{address-trans-cong-hpca2017}
Y.~{Hao} \emph{et~al.}, ``{Supporting Address Translation for
  Accelerator-Centric Architectures},'' in \emph{HPCA}, 2017.

\end{thebibliography}

\end{document}